\newcommand{\be}{\begin{eqnarray}}
\newcommand{\ee}{\end{eqnarray}}
\def\kms{km\thinspace s$^{-1}$\ }
\def\ms{m\thinspace s$^{-1}$\ }

\def\deg{$^\circ$}

\documentstyle[epsf]{mn}
\begin{document}
\title[The Spectral Variability of Aldebaran]
{On the Nature of the Radial Velocity Variability of Aldebaran:
A Search for Spectral Line Bisector Variations}

\author[Hatzes and Cochran]
{Artie P. Hatzes and William D. Cochran \\ 
McDonald Observatory, University of Texas at Austin, Austin, TX 78712 USA}
\date{}

\maketitle

\begin{abstract}
The shape of the Ti~I 6303.8\,{\AA} spectral line of
Aldebaran as measured by the line bisector
was investigated using  high signal-to-noise, high
resolution data. The goal of this study was  to understand
the nature of the 643-day period in the radial velocity
for this star reported by Hatzes and Cochran.
Variations in the line bisector with the radial velocity period
would provide strong evidence in support of rotational modulation
or stellar pulsations as the cause of the 643-day period.
A lack of any bisector variability at this period would
support the planet hypothesis.

Variations in the line asymmetries are found
with a period of 49.93 days. These variations are uncorrelated
with 643-day period found previously in the radial velocity
measurements. It is demonstrated that this 50-day period
is consistent with  an $m$ = 4 nonradial sectoral g-mode
oscillation. The lack of spectral variability with the
radial velocity period of 643 days may provide strong evidence
in support of the
hypothesis that this variability stems from the reflex
motion of the central star due to a planetary companion
having a mass of 11 Jupiter masses. However, this long-period
variability may still be due to a low order ($m$ = 2) pulsation mode
since these would cause bisector variations less than the 
error measurement.
\end{abstract}


\section{Introduction}

	Recent work has demonstrated that several K giant stars exhibit
low-amplitude, long-term radial velocity (RV) variations with periods of
several hundreds of days \nocite{mcc85} \nocite{irw89} 
\nocite{wal92} \nocite{hat93} (McClure et al. 1985; 
Irwin et al. 1989; Walker et al. 1992; Hatzes \& Cochran 1993). The nature
of these variations is presently unknown. Radial pulsations
can be excluded as a cause since the period
of the fundamental radial mode is expected to be about a week.
(These stars also show short-term
variability on timescales of a few days which are due to radial or nonradial
pulsations.) This leaves nonradial pulsations (NRP),
rotational modulation by surface features, or low-mass companions
as possible explanations for the long-period RV variations. K giant stars
have large radii and low projected rotational velocities so 
the period of rotation for these stars is expected to be several hundreds of days.
Rotational modulation thus seems to be a front-running hypothesis
for explaining the long-term variability. If the surface features
are related to stellar activity (spots, plage, etc.) then we should expect
to find variations in the equivalent widths of lines,
particularly ones formed in the chromospheric. Indeed, Lambert (1987) \nocite{lam87}
found variations in the He~I 10830\,{\AA} line in Arcturus    with 
the same period (233 days) that was later found in the RV variations.
Larson et al. (1993) found evidence for the 545 day RV period
in the equivalent width variations of Ca~II 8662\,{\AA} in $\beta$~Geminorum.
It thus seems that the long-period variability is consistent with
rotational modulation, at least for Arcturus     and $\beta$~Geminorum.

	The confirmation of rotational modulation as the source
of the RV variability 
has yet to be established for Aldebaran ($=$ $\alpha$~Tauri), a 
K giant with an RV period of 643 days 
and a  2-$K$ (peak-to-peak) amplitude of 400\,{\ms} (Hatzes \& Cochran 1993;
hereafter HC93). 
The interesting
aspect of this variability is that is seems  to have been present
and coherent (same amplitude and phase) in RV measurements
spanning over 12 years. If surface structure is responsible
for the RV variability of this star then it must be very long-lived,
which at first seems unlikely, but since  nothing is known about
surface structure on K giants this hypothesis cannot
be summarily rejected. The lifetime and coherency of the
long-period RV variations in Aldebaran    would normally argue
in favour of a low-mass companion. After all, one would expect changes in
the amplitude and phase of the variations 
if they were due to a surface 
structure that was evolving (changing shape, size, or
location on the stellar surface).  However, the fact that so many K giants show
long-period variability at about the same timescales makes us reluctant
to embrace the low-mass companion hypothesis.

	In a recent study of Aldebaran, Larson (1996)
found no evidence of the 643-day period in the equivalent width
measurements of the Ca~II 8662\,{\AA}, which is
an indicator of chromospheric activity. 
This casts some doubt on rotational
modulation as a possible explanation for the RV variability.
However, the surface features on Aldebaran    may be 
different from those which normally constitute stellar active
regions (spots, plage, faculae) 
and as such these may show no modulation in chromospheric
lines. One thing is certain though,
for a surface feature, regardless of its nature,
to produce RV shifts in the stellar lines it would have to alter
the line shapes. Consequently the RV variations should be
accompanied by spectral variability and the detection of these
would provide strong confirmation of 
 the rotational modulation hypothesis. Of 
the current explanations for the RV variability of Aldebaran, 
a low-mass
companion  seems to be the only one which should not produce changes in the
spectral line shapes. Both surface features (e.g. Toner \& Gray 1988;
Dempsey et al. 1992)
and nonradial pulsations  \cite{hat96} should produce
spectral line variations. A lack of such
variability would provide more support for the planetary companion
hypothesis.
	
	Aldebaran is a slowly rotating star so the best method
for discerning changes in the spectral line shapes  is via
line bisectors. Bisectors (the locus of the midpoints of 
horizontal line segments spanning the line width)
have proved to be a powerful tool for studying subtle
changes in the line shape due to the convection pattern on the
star(Gray 1982; Dravins 1987) or cool surface features
(Toner \& Gray 1988; Dempsey et al. 1992).  Recently, a study of the spectral
line bisectors of 51 Peg was recently used to  cast serious doubt on the
validity of the planet
hypothesis for this star (Gray 1997).
 	
	The objectives of this paper are simply 
to search for line bisector variability in Aldbaran. Such variability 
with the RV period would immediately eliminate the planet hypothesis as a cause
of this period. A lack of bisector variability would exclude rotational
modulation and possibly pulsations as viable hypotheses. The
organisation of this paper is as follows. The 
data are described in Section \S2 and the analysis of the spectral
line bisectors is presented in Section \S3 and it is demonstrated that
bisector variability is indeed present, but at a period unrelated
to the RV period. In Section \S4 the nature of the 
bisector variability is examined.
Finally, in Section \S5 the implications of the 
bisector variability on the interpretation
of the 643-day period is in examined.

\section{Data Acquisition}

	For the past several years at McDonald Observatory we have
been using precise stellar radial velocity measurements
($\sigma$ = 10--20 {\ms}) to study a variety of phenomena in stars,
including K giant variability. In the early years of this programme
the telluric O$_2$ lines were used as a wavelength reference for measuring
stellar radial velocity shifts. This technique was first
proposed by the Griffin \& Griffin  (1973) and we have confirmed their
claim
that it is capable of measuring stellar radial velocities to a precision
of 15\,{\ms}. These data are an ideal set for searching for
spectral variability as they were taken at high spectral
resolution,  with high signal-to-noise  ratio,  and 
only a few of the spectral lines are blended with telluric O$_2$.
There is the added advantage that the stellar
radial velocity and line shape measurements were made from the same
data set.

\begin{figure}
\epsfxsize=9truecm
\epsffile{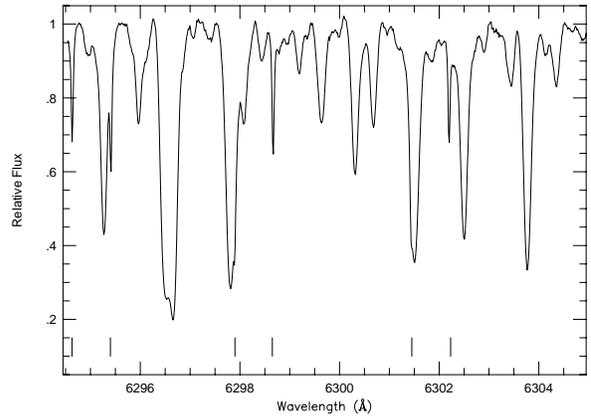}
\caption{The spectral region of Aldebaran    around 6300\,{\AA}.}
\label{fig:spectrum}
\end{figure}

	The observations were acquired at 
McDonald Observatory using the coud{\'e} focus of the 2.7-m 
telescope. An echelle grating was used in single pass along with 
a Texas Instruments 800$\times$800 3-phase CCD. The wavelength
coverage was 11.6\,{\AA} 
centred on 6300\,{\AA} at a spectral resolution
of  0.036\,{\AA} (130 $\mu$m slit which subtends 2 pixels on the 
CCD). An interference
filter was used to isolate the appropriate order from the echelle grating.
Figure~\ref{fig:spectrum} shows a typical spectrum of Aldebaran    taken with
this setup. The horizontal lines mark the location of the
telluric O$_2$ lines. (The wavelength scale of this spectrum
has been corrected for both the Earth's orbital motion
and the overall radial velocity motion of the star. Although
this places the stellar lines at the true  rest wavelength,
the telluric O$_2$ lines appear shifted from their proper
wavelength position.) Typical signal-to-noise ratio per resolution 
element for a given  exposure is about 450.

\section{Results}
\subsection{Line bisector measurements}

	For line bisector work one would like
to use a relatively deep, unblended line. Unfortunately,
Aldebaran    is a cool star with a plethora of stellar
lines so blending is a problem. From Figure~\ref{fig:spectrum}
 it appears
that Fe~I 6301.5\,{\AA}, Fe~I  6302.5\,{\AA}, and Ti~I 6303.8\,{\AA}  satisfy
this requirement. Regrettably, the two iron lines are often 
blended with a telluric  O$_2$ line and this severely limited
the  size of the data set used for searching for variability.
Consequently the Ti~I 6303.8\,{\AA}  line was best suited for
computing spectral line bisectors. It is never blended with a 
telluric line and the stellar lines  to either side
are far enough from the Ti~I feature so as not to affect the
overall shape of its bisector, except near the continuum.
Since we are primarily concerned with  changes in the
bisector shape, the small effect these blends may have on the absolute
shape of the 6304\,{\AA}  bisector is not a concern.

\begin{figure}
\epsfxsize=9truecm
\epsffile{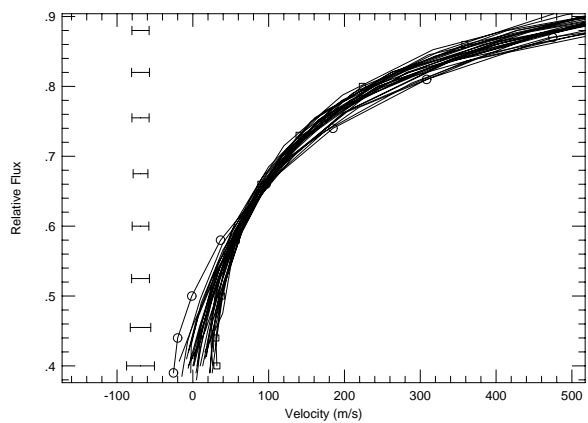}
\caption{The nightly mean bisectors of Ti 6303.8\,{\AA}. Circles and
squares mark two bisectors with extreme variations in shape.}
\label{fig:bisectors}
\end{figure}

	Figure~\ref{fig:bisectors} shows the spectral line bisectors of the 6304\,{\AA}  
line. On a given night typically 2--4 individual observations
were made of Aldebaran    and each bisector shown in Figure~\ref{fig:bisectors} 
represents a nightly mean. The horizontal lines to the left of
the bisector represents a typical error in the bisector
for a given flux level. These were computed assuming an average
of two bisectors. Frequently, 
more than two bisectors were used in computing the nightly
mean, so these bars  represent a ``worst case'' error. Although many
of the bisectors seem to fall on a  mean curve bounded by the
typical error, there appear to be instances of significant
variability. Two bisectors showing  extreme variability
are marked by  circles  and squares.

	To investigate in greater detail the bisector variability, the velocity
span of the individual (nightly averages) bisector was measured.
The velocity span, $S$,  is merely the difference in velocity between
two points lying on the bisector. In measuring this span it is important
to avoid regions of the bisector with the largest errors, namely
the core of the lines and the wings. The wings should also be
avoided due to line blending. The flux levels of 0.45 and 0.85
were chosen for the end points of the velocity span. Table~\ref{table1}
lists the velocity span measurements for the nightly bisector.
The typical error for a mean velocity span is about $\pm$25 {\ms}.

\begin{table}
\caption{Bisector Velocity Span Measurements}
\label{table1}
\begin{tabular}{lclc}
Julian Day    & {Span } & {Julian Day} &  {Span} \\
(-2440000)    & ({\ms}) & (-244000) & ({\ms}) \\
\hline
7429.9782&  370& 7785.9796 & 401  \\
7430.8255&  413& 7787.8543 & 343  \\
7430.9745&  379& 7813.9341 & 394  \\
7459.8904&  327& 7814.8974 & 367  \\
7459.9641&  332& 7879.8283 & 436  \\
7460.7790&  349& 7894.7800 & 331  \\
7495.8502&  350& 7895.7288 & 350  \\
7496.8148&  295& 7935.7002 & 424  \\
7516.7881&  370& 8145.9930 & 304  \\
7517.8108&  356& 8146.9962 & 279  \\
7551.7064&  313& 8176.9597 & 376  \\
7552.5996&  301& 8557.8879 & 322  \\
7611.6076&  349& 9260.9224 & 353  \\
\hline
\end{tabular}
\end{table}

\subsection{Period analysis}

\begin{figure}
\epsfxsize=9truecm
\epsffile{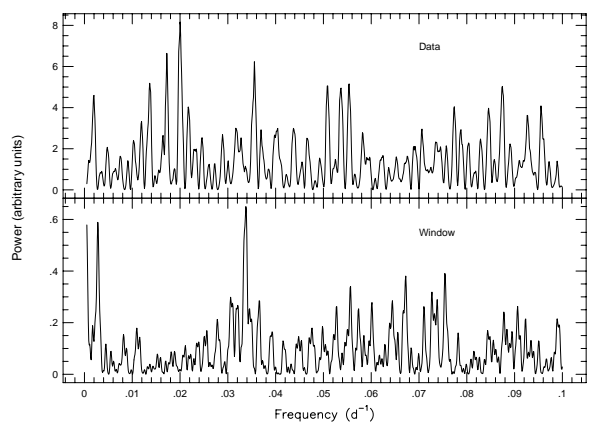}
\caption{(Top) The periodogram of the velocity span measurements.
(Bottom) The window function.}
\label{fig:ft}
\end{figure}

	A period analysis was performed on the velocity spans in
Table~/ref{table1}
using a Scargle-type periodogram \cite{sca82}. The results are shown
in the top of Figure~\ref{fig:ft}. The lower panel in the figure shows the 
window function. There is significant power at a period of 49.93 days and
the probability  that this peak arises from pure noise is 0.7\% 
using the equation in Scargle (1982). This false alarm
probability was also confirmed through numerical simulations.
First, a series
of 10,000 data sets consisting of random noise with $\sigma$ = 25 {\ms}
and sampled in the same manner as the data yielded peaks in the
periodograms that were larger than the data periodograms only
0.25\% of the time. Finally, the bisector span measurements
were randomly shuffled keeping the time values fixed. The fraction
of a large number (10,000) of these ``shuffled-data'' periodograms
also gave a measure of the false alarm probability. This value was
0.2\%.

\begin{figure}
\epsfxsize=9truecm
\epsffile{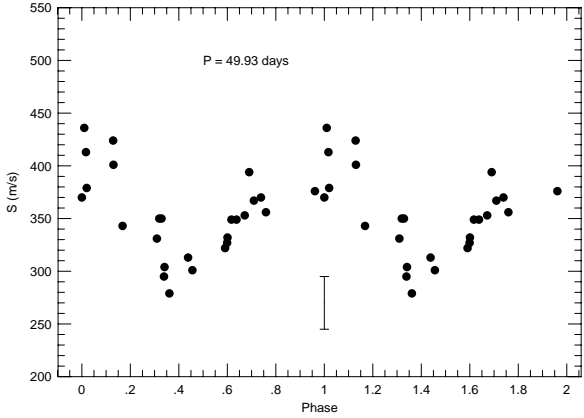}
\caption{The bisector span measurements phased to the period of 49.93 days.}
\label{fig:phase}
\end{figure}

	Figure~\ref{fig:phase} shows the velocity span of the bisectors phased
to the period of 49.93 days. The vertical line represents
a typical error for a span measurement. There are clear and convincing
periodic  variations with an amplitude at least twice that of the typical
error.

\subsection{Bisector variations of a constant star}

	It is important to establish that the bisector variations
found in Aldebaran    are indeed real and not due to some
systematic error. A prime candidate for such systematic
error is changes in the instrumental profile (IP) 
from run-to-run. These may cause slight changes in the shape
of the line profile which would manifest itself as bisector
variability. Although the coud{\'e} spectrograph is very stable
and great care is taken to ensure a constant IP  (as a testament
to this stability is the fact that the precision of our RV measurements
is about 15 {\ms}; large variations in the IP  would result in a
poorer RV precision) one should check that there is no significant
variability in the  bisector of a constant star.
	
	Fortunately, as part of the programme, routine observations
were made of the Moon which we  use as an RV standard.
These lunar observations have the advantages  that 1) they are of
comparable $S/N$ as the Aldebaran    data, 2) the Sun is known to
be constant (at least to the level that we are interested in),
and 3) the observations were taken on the same nights as the
Aldebaran    measurements. Because of the last point, any variations
in the spectral line shapes that are due to changes in the IP 
from night-to-night would be apparent in bisector measurements
for the Moon. Unfortunately, the Ti 6304\,{\AA} line  in the solar
spectrum is rather weak. However, two  strong, apparently
blend-free lines seemed to be suitable for bisector measurements:
the Fe~I 6302\,{\AA} line and the
Fe~I 6298\,{\AA} line. (The RV variations of the Moon were such
that the Fe~I 6302\,{\AA} line was always clear of the nearby
telluric line; this was not the case for Aldebaran.)
	
\begin{figure}
\epsfxsize=9truecm
\epsffile{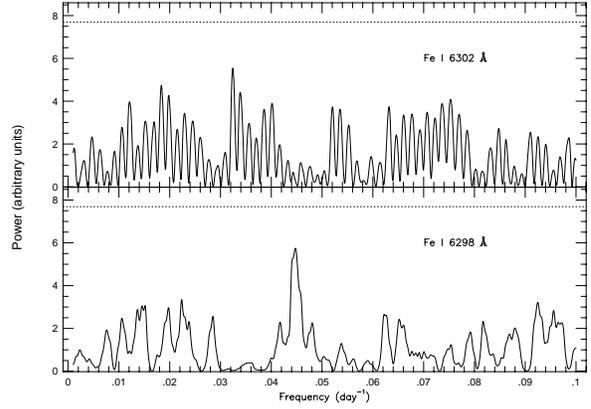}
\caption{Periodogram of the velocity span for the bisectors
of the Fe~I 6302\,{\AA} line (top) and the Fe~I 6298\,{\AA}
measured from spectra of the Moon.}
\label{fig:moon}
\end{figure}

	Figure~\ref{fig:moon} shows the periodogram for the bisector
span measurements of the Fe~I 6302\,{\AA} line (top panel) and
 the Fe~I 6298\,{\AA} line (bottom panel). The dashed line represents a false
alarm probability of 1\%. There are no significant peaks
at a period of 50 days, nor is any peak significant below a
false alarm probability of about 10\%.

\subsection{Stability of the instrumental profile}

\begin{figure}
\epsfxsize=9truecm
\epsffile{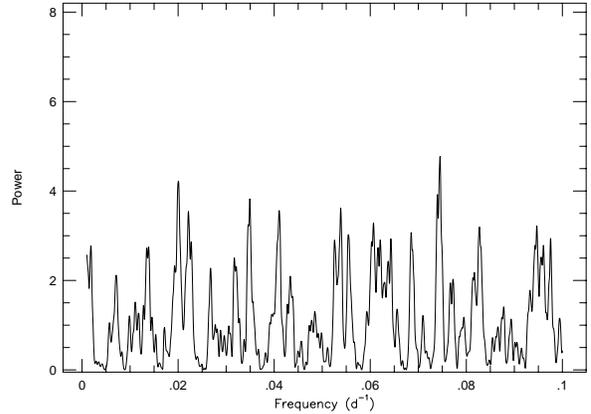}
\caption{Periodogram of the FWHM of a thorium emission
line nearest in wavelength to the Ti~I 6304\,{\AA} feature.}
\label{fig:thar}
\end{figure}

	As  a final test of the stability of the instrumental
profile, the full-width at half maximum (FWHM) of a thorium
emission line that was nearest in wavelength to the 
Ti 6304\,{\AA} feature was measured using spectra of calibration lamps
taken on the same nights as the data. The rms variations of the
FWHM was about 5\%. Figure~\ref{fig:thar}  shows the periodogram of the
measured FWHM for the thorium feature. Once again there
is no significant power near the period of 50 days ($\nu$ = 0.02
d$^{-1}$).

\section{On the Nature of the 50-day Bisector Variability}

	The residual RV data after removal of the 643-day period
shows no obvious evidence for the presence of a 50-day period (see
Section \S4). Therefore,
any mechanism for producing the line bisector 
variations must do so 	 without introducing 
significant RV variability. 
Hatzes (1996) examined the spectral and RV variations
expected for nonradial sectoral ($l$ = ${\pm}m$, where
$l$ and $m$ are quantum numbers of spherical harmonics)
pulsations in stars whose spectral lines
are not appreciably broadened by stellar rotation. These calculations
showed that the largest RV amplitude was  produced by the
low order $m$ = 2, 3 modes, but the largest change in the 
line bisectors occurred for the intermediate $m$ = 4 or 5 modes. 
Particularly for g-modes the magnitude of the bisector  variations
peaked sharply at $m$ = 4-5 and 
declined toward lower and higher $m$-number.

	To investigate whether nonradial pulsations can account
for the spectral variations of Aldebaran,
theoretical spectral line profiles from a nonradially
pulsating star were calculated using a disk integration on 
a synthetic star that was divided into
120$\times$120 cells. 
Local line profiles for each cell consisted of the Ti~I 6304\,{\AA}
generated using 20 limb angles and a model atmosphere
for a K1~III star
tabulated in Bell et al. (1976). 
A stellar inclination of 80{\deg} and 
a projected rotational 
velocity of 3 {\kms} were  adopted in the modeling.
The latter is comparable to the projected rotational velocity
measured for this star (HC93).
A macroturbulent velocity of 3 {\kms} was also included in the
calculation.
The dispersion and resolution for this synthetic
data was the same as the actual data. 
Only sectoral modes  ($l$ = $m$)  were 
calculated as these should produce the largest RV and spectral 
variations. 

In each stellar cell the combined Doppler shift
due  to 
stellar rotation, the macroturbulent
velocity, and the pulsational velocity  was applied to a local 
line profile prior to computing the integrated
profile. The contribution, in spherical coordinates,
 due to nonradial pulsations, 
of nonradial pulsations with an amplitude $V_p$ is given by
\begin{equation}
	V_r = V_p P_{\ell}^m(cos\, \theta)e^{{\it i}m\phi}
\end{equation}
\begin{equation}
	V_{\theta} = kV_p {d \over d{\theta}}
P_{\ell}^m(cos\,\theta)e^{{\it i}m\phi}
\end{equation}
\begin{equation}
	V_\phi = kV_p {P_{\ell}^m(cos\, \phi) \over sin{\theta}}
{{d \over d{\phi}}e^{{\it i}m\phi}}
\end{equation}
where $P_{\ell}^m$ is the associated Legendre function and
$k$ is a constant of proportionality between the horizontal
and vertical velocities. This horizontal scaler is not a free
parameter but is related to the frequency of oscillations
by:
\begin{equation}
{k =}{ {GM} \over {\sigma^2 R^3}}
\end{equation}
where $M$ and $R$ are the stellar mass and radius and
$\sigma$ is the pulsational frequency. For p-modes
$k$ is less than one, but is greater than unity 
for g-modes. In HC93  a 
mass of 2.5 $M_\odot$ and a radius of 42 $R_\odot$ were
used to calculate the period of the fundamental radial 
mode. These values along with a pulsation period of
50 days yield $k$ $\approx$ 6.

\begin{figure}
\epsfxsize=9truecm
\epsffile{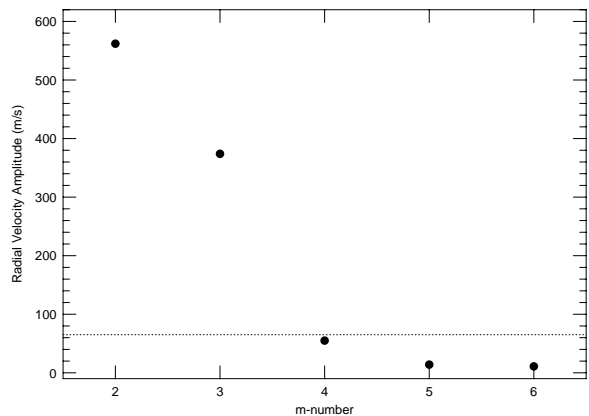}
\caption{ The predicted radial velocity amplitude from nonradial
pulsations which also produce a change in the line bisector
velocity span of 100 {\ms}.}
\label{fig:rvamp}
\end{figure}

	Figure~\ref{fig:rvamp} 
 shows the predicted radial velocity variation, as
a function of $m$-number for nonradial pulsations that will  also
produce a peak-to-peak change in the velocity span of the bisector of 100 {\ms},
as required by Figure~\ref{fig:phase}.
The velocity span points for simulations were taken at the same
flux level as the Aldebaran    measurements.
Since the phase diagram of the velocity
residuals 
shows no clear sinusoidal variations at a  50-day
period, we have adopted the rms scatter of velocity residuals
(after removal of the 643-day component)
as an upper limit to RV-amplitude for  such a period.
This rms scatter of about 65 {\ms}, which  is marked
with a horizontal dashed line in Figure~\ref{fig:rvamp}, 
is significantly
larger than the error of the radial velocity measurement which
is about 15--20 {\ms}. 
It is not known whether the 
scatter in the RV residuals is due to the 50-day component of
the variability of Aldebaran    or other short-term variability,
but presently we shall assume the former.
[Larson (1996) found evidence for a 1.8-day in her RV measurements
for Aldebaran    which may contribute to the scatter of our RV
residuals. There is no evidence for this period in either our
bisector or RV measurements, but this is most
 likely due to the sparse sampling
of our data.]
It seems that the observed changes in the spectral line bisectors
and the rms scatter of the radial velocity residuals are consistent
with an $m$ = 4 nonradial sectoral mode. Clearly, for an $m$ = 2
or 3 mode to produce the observed bisector variations this would
also produce a large radial velocity which should be seen in the
RV residuals. In fact, the amplitude of such a mode would be larger than 
the amplitude of the 643-day period, so 50 days should be the
dominant period in the RV variations and this is not the case.
Likewise the $m$ = 5 or 6 modes can be eliminated since these 
would cause too little variations in the RV  residuals. In fact,
if there were no other short-term variability present, the rms scatter
of the RV residuals would be no more 
than  about 25 {\ms} (this results from adding in quadrature
the error of the RV measurement with the RV amplitude of the 
nonradial pulsations). However, the presence of high-order
($m >$ 4) cannot be eliminated purely on the basis of the
rms scatter of the RV residuals. It may be that a large portion
of this scatter may be due to other short-term variability.
If so, then low-amplitude RV variations ($\sim$ 10 {\ms}) due
to high-order modes may be present and could also account for the
bisector variations.

\section{On the Nature of the 643-day RV period}

	The analysis of the bisector data revealed, to our surprise,
no evidence for variability with the 643-day period. What are the implications
for the various hypotheses that have been invoked to explain the RV variability?
Before this can be answered we must be certain that 1) the 50-day period
is not merely an alias of the 643-day period and 2) that a 643-day
period is not hidden somehow in the bisector variations.

\subsection{Where is the 643-day Period?}

	The periodogram in Figure~\ref{fig:ft} shows  no evidence
for the 643-day period found in the radial velocity measurements of HC93.
It is impossible for the long period to have been 
absent at the time of the bisector measurements since both these and 
the RV values were determined from the same data set.
The closest peak is one at 550 days,
a feature that is most likely associated with 
a feature in the window function. 

Monte Carlo simulations were performed to investigate whether
aliasing or noise can cause a 643-day period to appear
as a peak at  50 days in the periodogram. Synthetic
data using  a sine waves with a period of 643 days
and an amplitude of 50 {\ms} (the amplitude
of the span variations) were generated. These sine
waves were sampled in the same manner as the data and random 
noise at a level of 
$\sigma$ = 25 {\ms} was also added. A series of 1000 synthetic
data sets
were generated and the periodograms calculated.  In all but 18 cases
the highest peak of the periodogram coincided with the input period.
Of those 18 discrepant periods only one (at 60 days) was marginally close to
the 50-day period of the bisector periodogram. (The other
false periods occurred at 28 days, 20 days, 14 days, and 11 days.)
We thus estimate that the chance of the 50-day period in the bisector
measurements being caused by noise and alias effects of a true
643-day period is no more than 0.001.
We are convinced that the 643-day period is not present in the 
bisector velocity span measurements.

\begin{figure}
\epsfxsize=9truecm
\epsffile{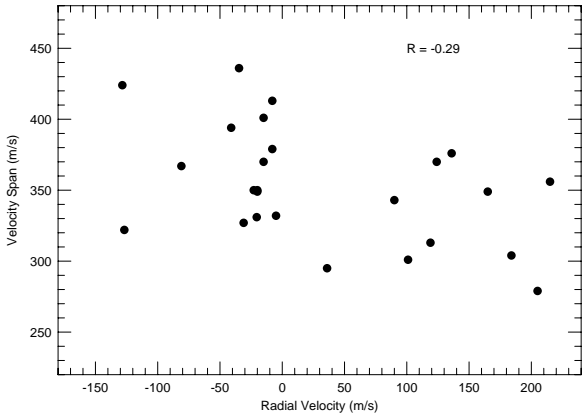}
\caption{The bisector velocity span measurements as a function of
radial velocity.}
\label{fig:correlate}
\end{figure}

\begin{figure}
\epsfxsize=8truecm
\epsffile{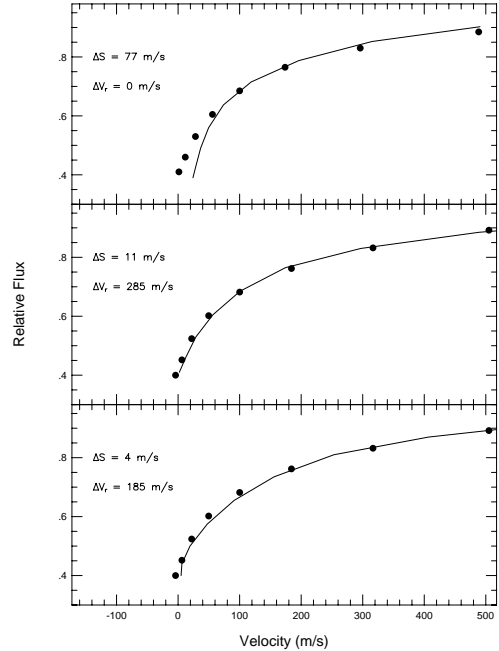}
\caption{(Top) Two line bisectors showing significantly different
shapes but which are not accompanied by measurable change in the
radial velocity. (Lower two panels). Two bisectors that have
large differences in the RV measure, but no measurable changes
in the line bisector shape.}
\label{fig:compare}
\end{figure}

	Further evidence that the 50-day bisector period is uncorrelated to 
the RV variations is shown in Figure~\ref{fig:correlate}
 which shows the velocity  span
of the bisectors plotted as a function of the RV measurements.
Although
there  seems to be a slight trend of decreasing velocity span with increasing
RV velocity, the correlation coefficient, $R$ = $-0.29$, indicates that the
two measurements are uncorrelated. More convincing evidence that
the bisector span variations are unrelated to the RV variability
is provided by a closer examination of some individual bisectors.
The top panel of Figure~\ref{fig:compare} 
shows two line bisectors with the same
radial velocity but whose velocity span
differ by almost 80 {\ms}.  The lower two panels each show two bisectors
whose velocity span differ only by 11 and 4 {\ms}, respectively, 
but whose radial velocity difference is 285 and 185 {\ms}, respectively.
This figure demonstrates that large bisector variations are not always
accompanied by large changes in the radial velocity. There are other instances
when  changes in the RV measures produces 
no measurable change in the bisector span.

\begin{figure}
\epsfxsize=9truecm
\epsffile{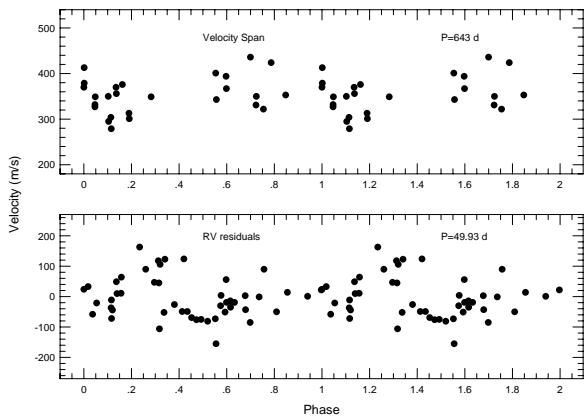}
\caption{(Top) The bisector velocity span measurements phased
to the 643-day RV period. (Bottom) The residual RV measurements
(after subtracting the 643-day component) phased to the bisector
period of 49.93-days.}
\label{fig:phase2}
\end{figure}

	Finally, the top panel of Figure~\ref{fig:phase2}
 shows the velocity span
measurements phased to the RV period of 643 days. The lower 
panel shows the radial velocity residuals (from HC93)
after subtraction of the
643-day component, phased to the bisector period of 49.93 days.
There is no strong evidence (as confirmed by periodogram analyses)
of the 643-day period in the bisector span measurements or
of a 49.93-day variation in the residual radial velocity measurements.

	The absence of the 643-day RV period in the line 
bisector variations places better constraints on the nature
of the long-term RV variability. The fact that the RV and line bisector
variations are uncorrelated firmly excludes rotational modulation
as a cause of the 643-day period. This leaves two hypotheses
for the  source of the RV variability: low-order ($m$ = 1-3)
nonradial pulsations or a planetary 
companion.

\subsection{Planetary companion}

	Table~\ref{table2} lists the implied orbital parameters for the
possible companion to Aldebaran.
(Note that the orbital solution found a slightly different period.
The RV period will still be referred to as the 643-day period which is the one
found via periodogram analysis). The fit of this orbit to the
RV measurements can be found in HC93. The implied
$m$ sin $i$ for the companion is 11 Jupiter masses.

\begin{table}
\caption{Orbital solution to the RV measurements of Aldebaran}
\label{table2}
\begin{tabular}{lcc}
{Parameter} & {Value} & Uncertainty \\
\hline
Orbital Period $P$ (days) &  653.8 & 10.1 \\
Velocity Semi-amplitude $K$ (m\,s$^{-1}$) & 147.2 & 8.1 \\
Eccentricity $e$ & 0.182 & 0.065 \\
Longitude of Periastron $\omega$ (degrees) &290.2 & 4.9 \\
Periastron Date $T_0$ (Julian Date) & 2447625.4 & 7.8 \\
\hline
\end{tabular}
\end{table}

	There are a number of arguments in support of the
planet hypothesis for the RV  period, besides the lack
of bisector variability at this period.
First, as established in HC93, the long-term RV period in 
Aldebaran    has been present at the same amplitude and phase
for at least 12 years. It is difficult to understand
why a single, long-period mode could be so long-lived.
Indeed, one K giant, Arcturus, shows a number of presumably nonradial
modes, not all of which seem to be present at any given time
(Smith et al. 1987; Belmonte et al. 1990; Hatzes \& Cochran 1994).
Also, if the 643-day period is due to low-order
modes then this implies that two nonradial modes have
periods that differ by more than an order of magnitude which also
seems unlikely. However, long-term stable pulsations certainly  exist 
in Cepheid variables and it may well be that different pulsation modes
have periods that can differ by more than a factor of ten. Also,
the two periods may result from two distinct pulsation modes. For instance
the 50-day period may result from nonradial $g$-modes, whereas the
643-day period may result from Rossby ($r$-modes). Theoretical
studies are needed to resolve these issues in particular whether
$r$-modes can produce significant RV variability.

\subsection{Nonradial pulsations}
	
	It is possible for  a low order ($m$ = 2) nonradial 
pulsation mode to
cause the RV variations
since the amplitude of the bisector velocity span 
variations should be significantly less than those produced
by an $m$ = 4 mode. But are these small enough to have
been missed in the bisector analysis? The RV amplitude of 150 {\ms}
is reproduced by an $m$ = 2 mode with an amplitude of only 0.2 {\ms}
(using $k$ = 1000 determined from Eq. 4). This results in variations
in the bisector velocity span with an amplitude of 15-20 {\ms} which
is less than the error of the measurement ($\sigma$ = 25 {\ms}).
One could be argue that 
if this additional bisector variability were present then it
should contribute to the error of the bisector measurement.
It is true that rms scatter of the bisector measurements in the
phase curve (Fig.~\ref{fig:phase}) is near that expected from photon
statistics which tentatively supports the lack of other sources for bisector
variability. However,  the number of data measurements are too few and the exact
error of the bisector measurements  too ill-determined to say
anything conclusive. (For example bisector error  of
20 {\ms}  added in quadrature to a variation with a 15 {\ms}  can
reproduce the scatter about the mean curve in Fig.~\ref{fig:phase}).

	The shape of the RV curve may provide further constraints
on the nature of the long period variations and this shape is best
quantified by the  eccentricity  of the ``orbital'' solution.
The eccentricity of the orbit is $e$ = 0.147 $\pm$0.008.
The predicted RV curve  from an $m$ = 2 pulsation mode (true 
for all sectoral modes) is very nearly sinusoidal (see Gray \& Hatzes 1997)
with 
an eccentricity of only 0.002. This would argue against pulsations as
a mechanism for producing the RV variations, but again this is not
conclusive for a number of reasons. 
The pulsational period is quite large compared to that
of the fundamental radial mode and there can be temperature
variations or non-linear affects.
The long period also implies a high radial order ($n$ $\approx$ 100) for the
pulsations which translates into a rather  short vertical wavelength. Consequently,
the vertical velocity structure may be changing rapidly through the line forming
region which may affect the source function.
All of these mechanisms  may affect the shape 
of the integrated
RV curve resulting in a slightly eccentric ``orbit.''
The predicted RV curve from the pure kinematic model
which does not take into account any temperature variations or  any
vertical velocity structure to the pulsations may thus be too simplified
a treatment for a complicated physical process. Finally, there is no
reason to expect only sectoral modes. Other modes, or combination
of modes may be present and these may produce an RV curves that departs
from a pure sine wave.

\section{Discussion}
	
	The 50 day period found in the line bisector variations of
Aldebaran is most likely associated with a nonradial pulsation mode.
The long period of the variations implies a $g$-mode oscillation.
Although this variation can be modeled with an $m$ = 4 nonradial mode,
we cannot exclude the possibility that this arises from a higher order
mode, possibly even one that is non-sectoral ($l$ $\ne$ $m$).
The bisector period 
found in Aldebaran may be related to one in Arcturus.
That star showed 
a 46 day period with an amplitude $=$ 50 {\ms}  in the  RV measurements
after subtracting the dominant 233 day component (HC93).
Possibly these $\approx$ 50 day periods represent an intermediate $g$-mode
between the very short ($\approx$ days) $p$-mode oscillations and the
very long
($\approx$ years) pulsation modes.

	The lack of line bisector variability in Aldebaran
with the 643-day RV period seems to  support the hypothesis
that these variations are due to a planetary companion. However,
the presence of a low order $m$ = 2 nonradial mode can also
account for  RV variations and apparent lack of bisector variability.
Also, the fact that two periods have now been identified in Aldebaran
would argue in favour of the pulsation hypothesis for both periods.
Clearly, other types of measurements are needed before we can
distinguish between the two hypotheses.

	As previously mentioned, if the 643-day period is indeed due to a nonradial g-mode, then
the vertical wavelength should be quite short. If the vertical velocity 
structure of the atmosphere is indeed changing rapidly, then this might
manifest itself in the RV behaviour of different spectral lines. For instance,
strong lines are formed, on average, higher up in the stellar atmosphere
than  weak lines. An investigation of the  radial velocity as a function of
line strength may reveal differences in the amplitude and phase between
strong and weak lines which would provide confirmation of the pulsation
hypothesis.
Unfortunately, there are too few telluric O$_2$ lines (used as the wavelength
reference) to do a line-by-line analysis of the radial velocity of Aldebaran
with the data used in this study.

	Photometric measurements may also help us chose  between a planet
or pulsations, although these may be difficult to make. The analysis
of Buta \& Smith (1979) can be used to estimate the expected photometric 
variations for Aldebaran. Geometric effects will only be considered because,
as pointed out by Buta \& Smith, temperature variations for long period
pulsations are difficult to predict as these can be in-phase or
out-of-phase
with the geometric variations. Furthermore, the predicted photometric amplitudes
for the temperature variations can be orders of magnitude too large due to 
nonadiabatic effects. The predicted photometric amplitude from geometric 
effects is only about
$\Delta$V $\approx$ 0.1 mmag for the 643-day period
(assuming an $m$ = 2 mode). Considering the brightness of Aldebaran (and
the difficulty of finding a suitable comparison star) this may be an impossible
measurement.

	Photometry may be better able to confirm the 50-day bisector period.
If this is indeed due to an $m$ = 4, then the predicted pulsational amplitude
is about 40 {\ms} which would produce a photometric amplitude
(from geometrical effects)  of about $\Delta$V $\approx$ 0.6 mmag which
is comparable to the precision of
the best ground-based photometric measurements (Henry et al. 1996)

	In summary we have found evidence for changes in the
spectral line shape of the Ti~I 6303.8\,{\AA} feature in Aldebaran    with a period
of 49.93 days. These variations are uncorrelated with the 643-day
period found previously in the radial velocity measurements for
this star. The 50-day period most likely arises from an 
$m$ = 4 mode, although higher order modes cannot be 
excluded. The lack of spectral variations with a 643-day period
excludes rotational modulation as a source of the RV
variations. At the present time the planet hypothesis
seems to be the simplest one which can explain the 643-day RV period 
as well as the 
lack of bisector variability with that period.  However, due to the fact that so
many K giants exhibit multi-periodic RV variations makes us somewhat reluctant
to declare with certainty that Aldebaran is another star with a confirmed 
planetary companion.
Ancillary measurements are needed
to decide between these competing hypotheses.

\section{Acknowledgments}
	The authors wish to thank the A. Larson for communicating
her result on the lack of Ca~II equivalent width variations
in Aldebaran. This study was largely inspired by that result.	
This work was supported by NASA through grants NAGW-3990 and NAG5-4384.


\begin{thebibliography}{}
\bibitem[Bell et al. 1976]{bel76}
Bell, R.~A., Eriksson, K., Gustafsson, B., \& Nordlund, {AA}
1976, A\&AS, 23, 37.
\bibitem[Belmonte et al. 1990]{bel90}
Belmonte, J.~A., Jones, A.~R., Pall\'e, P.~L., \& Cortes, T.~R., 1990,
ApJ, 358, 595.
\bibitem[Buta \& Smith 1979]{but79} Buta, R.~J. \& Smith, M.~A. 1979, ApJ, 213.
\bibitem[Dempsey et al. 1992]{dem92} Dempsey, R.~C., Bopp, B.~W.,
Strassmeier, K.~G., Granados, A.~F., Henry, G.~W., and Hall, D.~S. 1992,
ApJ, 392, 187.
\bibitem[Dravins 1987]{dra87} Dravins, D. 1987, A\&A, 172, 200.
\bibitem[Gray 1982]{gra82} Gray, D.~F. 1982, ApJ, 255, 200.
\bibitem[Gray 1997]{gra97} Gray, D.~F. 1997, Nature, 385, 795.
\bibitem[Gray \& Hatzes]{gra91b} Gray, D.~F. \& Hatzes, A.~P. 1997,
ApJ, In press.
\bibitem[Griffin \& Griffin 1973]{gri73} Griffin, R. \& Griffin, R. 1973,
MNRAS, 162, 243.
\bibitem[Hatzes 1996]{hat96} Hatzes, A.~P. 1996, PASP, 109, 839.
\bibitem[Hatzes \& Cochran 1993]{hat93} Hatzes, A.~P, \& Cochran, W.~D. 1993,
ApJ, 413, 339 (HC93).
\bibitem[Hatzes \& Cochran 1994]{hat94} Hatzes, A.~P. \& Cochran W.~D. 1994, ApJ, 432, 763.
ApJ, 432, 763.
\bibitem[Henry et. al. 1997]{hen97} Henry, G., Baliunas, S.~L.,
Donahue, R.~A., Soon, W.~H., \& Saar, S.~H. 1997, ApJ, 474, 503.
\bibitem[Irwin et al. 1989]{irw89} Irwin, A.~W., Campbell, B.,
Morbey, C.~L., Walker, G.~A.~H., Yang, S. 1989, PASP, 101, 147.
\bibitem[Lambert 1987]{lam87} Lambert, D.~L. 1987, ApJS, 65, 255.
\bibitem[Larson 1996]{lar96} Larson, A.~M. 1996, Ph.D. Thesis,
University of Victoria.
\bibitem[Larson et al. 1993]{lar93} Larson, A.~M., Irwin, A.~W.,
Yang, S.~L.~S., Goodenough, C., Walker, G.~A.~H., Walker, A.~R., \&
Bohlender, D.~A., PASP, 1993, 105, 332.
\bibitem[McClure et al. 1985]{mcc85} McClure, R.~R., Griffin, R.~F., Fletcher, J.~M., Harris,
H.~C., \& Mayor, M. 1985, PASP, 97, 740.
\bibitem[Scargle 1982]{sca82} Scargle, J.~D. 1982, ApJ, 263, 835.
\bibitem[Smith et al. 1987]{smi87} Smith, P.~H., McMillan, R.~S., \&
Merline, W.~J. 1987, ApJL, 317, L79.
\bibitem[Toner \& Gray 1988]{ton88} Toner, C.~G. \& Gray, D.~F 1988,
ApJ, 334, 1008.
\bibitem[Walker et al. 1992]{wal92} Walker, G.~A.~H., Bohlender, D.~A.,
Walker, A.~R., Irwin, A.~W., Yang, S.~L.~S., \& Larson, A. 1992, ApJL,
396, 91.
\end{thebibliography}
\end{document}